# SIMPLE METALS AT HIGH PRESSURE

OLGA DEGTYAREVA[*]

*Centre for Science at Extreme Conditions and School of Physics, the University of Edinburgh, the King's Buildings, Mayfield Road, Edinburgh EH93JZ, United Kingdom*

**Abstract.** In this lecture we review high-pressure phase transition sequences exhibited by simple elements, looking at the examples of the main group I, II, IV, V, and VI elements. General trends are established by analyzing the changes in coordination number on compression. Experimentally found phase transitions and crystal structures are discussed with a brief description of the present theoretical picture.

## 1. Introduction

Most of the metallic elements crystallize in the closely-packed atomic arrangements, body-centered cubic (bcc), face-centered cubic (fcc) and hexagonal close-packed (hcp) (*Young 1991*). The group IV, V, IV and VII semimetals and non-metals crystallize in open packed structures following the 8-N coordination rule. According to this rule, a structure is formed with a coordination number (i.e. the number of nearest neighbors) equal to 8-N, where N is the group number in the periodic table of elements. As a rule of thumb, most elements in the middle of the periodic table (the so-called transition metals), that crystallize in the closely-packed structures (see crosses in Figure 1), show no structural transitions or transition to other closely-packed structures. For example, Au transforms from fcc to hcp at pressures of 240 GPa and high temperatures (*Dubrovinsky et al. 2007*). Most elements from the left and from the right of the periodic table (so-called simple or *sp*-elements) undergo on compression a series of structural phase transitions that show diverse and sometimes unexpectedly complex behavior. The non-metallic elements from the group IV, V, and VI on the right of the periodic table having large initial atomic volume compared to the transition metals are strongly compressed under pressure, showing a transition to a metallic state with closely-packed structures (see naughts changing to crosses in Figures 1a and b). The group I and II elements (the alkali and alkali-earth elements) on the left of the periodic table crystallize in closely-packed structures, however under pressure show anomalous behavior transforming to open packed structures (see crosses changing to naughts in Figures 1a and b). In comparison to the transition metals, they have very large initial atomic volumes that decrease drastically on pressure increase while the elements undergo transitions in atomic and electronic structure.

---
[*]To whom correspondence should be addressed. Olga Degtyareva, Centre for Science at Extreme Conditions and School of Physics, the University of Edinburgh, the King's Buildings, Mayfield Road, Edinburgh EH93JZ, United Kingdom; e-mail: O.Degtyareva@ed.ac.uk

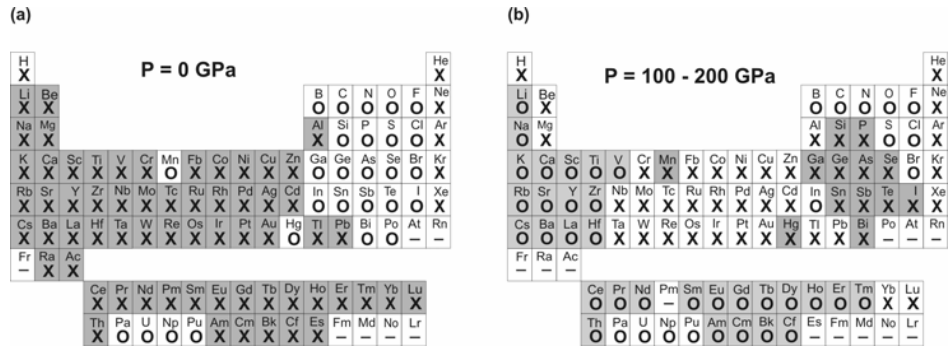

*Figure 1.* Periodic table with crystal structures of elements (a) at ambient pressure and (b) high pressure of 100-200 GPa (modified from *Young 1991*). Crosses denote the closely-packed structures, such as body-centered cubic (bcc), face-centered cubic (fcc) and hexagonal close-packed (hcp) structures. Naughts denote open packed and distorted structures. In (b), data are taken from the papers discussed in the text. In (a), the closely-packed structures (crosses) are highlighted, in (b), the changes from crosses to naughts and from naughts to crosses are highlighted

The high-pressure behavior of elements is described in several reviews, some of the recent being: (*Degtyareva 2006, McMahon and Nelmes 2006, Schwarz 2004, Tonkov and Ponyatovsky 2005*). In the present lecture we focus on the experimentally found transitions in atomic structure in the main group I, II, IV, V and VI elements with the emphasis on the most recent results reported in years 2006-2009. The lanthanide and actinide elements, although showing lots of changes from crosses to naughts in Figure 1, are not discussed in this lecture. An interested reader is referred to the review by (*Tonkov and Ponyatovsky 2005*) and the recent experimental reports on the phase transitions from close-packed to distorted structures in lanthanides (see e.g. *Cunningham et al. 2007, Errandonea et al. 2007, Pravica et al. 2007, Shen et al. 2007*).

## 2.   Phase transitions under high pressure

### 2.1.  GROUP-IV ELEMENTS

The semiconducting elements Si and Ge crystallize in a diamond-type structure with the coordination number equal 4, according to the 8-N rule. On pressure increase they transform to a β-Sn structure (white tin), while this transition occurs in Sn at ambient pressure on temperature increase (Table 1) (see (*Schwarz 2004* and references therein). On further pressure increase, Si and Ge transform to an orthorhombic structure with four atoms in the unit cell (Pearson notation, *Pearson 1967*) *oI*4, where "o" stands for orthorhombic, "I" for body-centered, "4" for the number of atoms in the unit cell). The *oI*4 phase transforms to a simple hexagonal structure (sh, Pearson notation *hP*1). All these structures are very simple containing a small number of atoms in the unit cell. Then on further pressure increase, both Si and Ge transform to a crystal structure with an unexpectedly complex arrangement of atoms, with 16 atoms in an orthorhombic unit cell (Pearson notation *oC*16) occupying two different atomic sites 8*d* and 8*f* of the space group *Cmca*.

This phase has long been known to exist in Si, however it was solved only relatively recently: its solution is reported in the year 1999 (*Hanfland et al. 1999*); the same structure was reported for a high-pressure phase of Cs in 1998 (*Schwarz et al. 1998*). These experimental studies took advantage of the new 2D image plate techniques that became available at the beginning of 90's. These publications reported the first structure solution of a complex phase in a simple element and marked a new era of pressure-induced complexity of simple elements. Reports of many other complex structures started to follow. Apart from Si and Ge, the *oC*16 structure has been reported for the alkali metals Cs (as mentioned above) and Rb (section 2.4), a high-pressure high-temperature modification of Bi, Bi-IV, as well as for Bi-In, Bi-Pb and Bi-Sn alloys, where a site-ordering was observed (see (*Degtyareva et al. 2003*) and references therein). The fact that this complex structure appears under pressure in alkali metals as well as in polyvalent metals and alloys deserves special attention, as discussed in (*Degtyareva 2006*).

TABLE 1. Sequences of structural transformations on pressure increase for group IV elements. The numbers above the arrows show the transition pressures in GPa. The crystal structures of the phases are denoted with their Pearson symbols, apart from the most common metallic structures bcc, fcc and hcp

| | | | | | | | |
|---|---|---|---|---|---|---|---|
| **Si** | $cF8 \rightarrow$ Diamond | $\overset{12}{\beta\text{-Sn, } tI4} \rightarrow$ white tin | $\overset{13}{oI4} \rightarrow$ | $\overset{16}{hP1} \rightarrow$ | $\overset{38}{oC16} \rightarrow$ | $\overset{42}{\text{hcp}} \rightarrow$ | $\overset{80}{\text{fcc}} < 250$ GPa |
| **Ge** | $cF8 \rightarrow$ Diamond | $\overset{11}{\beta\text{-Sn, } tI4} \rightarrow$ white tin | $\overset{75}{oI4} \rightarrow$ | $\overset{85}{hP1} \rightarrow$ | $\overset{102}{oC16} \rightarrow$ | $\overset{160}{\text{hcp}} < 180$ GPa | |
| **Sn** | $cF8 \rightarrow$ Diamond | $\overset{0}{\beta\text{-Sn, } tI4} \rightarrow$ white tin | $\overset{8}{tI2} \rightarrow$ | $\overset{45}{\text{bcc}} < 120$ GPa | | | |
| **Pb** | fcc $\overset{13}{\rightarrow}$ | hcp $\overset{110}{\rightarrow}$ | bcc $< 270$ GPa | | | | |

On further pressure increase, the $oC16$ structure transforms to hcp in both Si and Ge, and further to fcc in Si. Thus, a complete phase transition sequence is observed from an open-packed semiconducting structure of a diamond-type to a metallic close-packed structures hcp and fcc. The coordination number increases gradually with pressure increase from 4 in the diamond-type structure to 6 in $\beta$-Sn, to 8 in simple hexagonal, to 10-11 in $oC16$ and finally to 12 in hcp and fcc. The packing density increases from 0.34 in diamond structure to 0.55 in $\beta$-Sn, to 0.6 in simple hexagonal to 0.74 in hcp and fcc, the highest packing density for regular spheres. The complex $oC16$ structure plays an important role in this transition sequence providing an atomic arrangement with coordination number 10-11, intermediate between those of open-packed and close-packed structures. The complex atomic arrangement of the $oC16$ phase, a nearly-free electron metal in the case of Si and Ge, provides a reduction of the electronic energy, which reduces the total structure energy (see for example *Degtyareva 2000*)

## 2.2. GROUP-V ELEMENTS

### 2.2.1. *Arsenic, antimony and bismuth*

At ambient pressure, the group-V elements As, Sb and Bi crystallize in a rhombohedral As-type structure with two atoms in the unit cell (Pearson notation $hR2$), space group $R\text{-}3m$ (Figure 2, left). Under pressure these elements show a similar phase-transition sequence, as summarized in Table 2. Arsenic transforms from the $hR2$ phase to a simple cubic (sc) structure with one atom per unit cell (Pearson notation $cP1$). The sc structure is stable up to 48 GPa, where it transforms to As-III, which has a monoclinic host-guest structure (*Degtyareva et al. 2004 a & b*). For Sb, the $hR2$ structure approaches the sc structure on compression, but before it is reached, Sb transform to a monoclinic host-guest structure at 8.0 GPa, followed by a tetragonal host-guest structure at 8.6 GPa (*Degtyareva et al. 2004 a & b, Schwarz et al. 2003*). For Bi, the $hR2$ structure also approaches sc under pressure but transforms at 2.55 GPa to Bi-II, which has a monoclinic structure with 4 atoms in the unit cell (Pearson notation $mC4$). This structure is unique amongst the elements but can be considered as a distorted sc structure. Bi-II has a very narrow stability range, and it transforms at 2.69 GPa to the tetragonal host-guest structure of Bi-III (*McMahon et al. 2000a*) (Figure 2, middle). At higher pressures, the host-guest structures of As, Sb and Bi all transform to a bcc phase (Figure 2, right). This is the current end member of the known high-pressure structural sequences of these elements (Table 2).

The host-guest structure of group-V elements comprises a body-centered tetragonal (monoclinic) host framework, in which Bi atoms stack in such a way as to create channels along the c-axis. Within these channels lie chains of other Bi atoms that form a body-centered tetragonal (monoclinic) guest structure. The host and guest structures are incommensurate with each other along their common $c$ axis, with $c_H/c_G \sim 1.31$. There is a non-integer number of atoms (10.62 in total) in the host unit cell. Both host and guest structures are modulated along the c-axis, as shown from the recent single-crystal study of the Bi-III structure (*McMahon et al. 2007a*).

A crystallographic description of such host-guest phases requires an application of a (3+1)-dimensional superspace formalism (see Section 3.1). The superspace group for the Bi-III structure is $\varGamma 4/mcm(00\gamma)0000$, where $\gamma$ is the (incommensurate) ratio of the $c$-axis lattice parameters of the host and guest structures, $c_H/c_G$, and $\varGamma$ is the centering (½ ½ ½ ½) in superspace.

The coordination number increases under pressure from 3 in the $hR2$ (As-type) structure to 6 in the simple cubic structure of As or 7 in the $mC4$ structure of Bi, to 9 to 10 for the host-guest structure and finally to 14 (=8+6) in the ultimate bcc structure. The complex atomic arrangement of the modulated host-guest structure (coordination number 9 for the host atoms and 10 for the guest atoms) plays a role of an intermediate step between the open packed semi-metallic As-type structure and metallic closely-packed bcc structure.

TABLE 2. Sequences of structural transformations on pressure increase for group-V elements. The numbers above the arrows show the transition pressures in GPa. . "H-g" stands for a host-guest structure, "inc. mod" stands for an incommensurately modulated structure

| **P** | $oC8$ $\rightarrow$ $hR2$ $\xrightarrow{5}$ $cP1$ $\xrightarrow{10}$ inc. mod. $\xrightarrow{103}$ $hP1$ $\xrightarrow{137}$ bcc $< 280$ GPa $\xrightarrow{260}$ |
| **As** | $hR2$ $\xrightarrow{25}$ $cP1$ $\xrightarrow{48}$ monocl. h-g $\xrightarrow{97}$ bcc $< 122$ GPa |
| **Sb** | $hR2$ $\xrightarrow{8.0}$ monocl. h-g $\xrightarrow{8.6}$ tetr. h-g $\xrightarrow{28}$ bcc $< 65$ GPa |
| **Bi** | $hR2$ $\xrightarrow{2.5}$ $mC4$ $\xrightarrow{2.7}$ tetr. h-g $\xrightarrow{7.7}$ bcc $< 220$ GPa |
|  | $\rightarrow oC16$ (>483K) $\rightarrow$ |

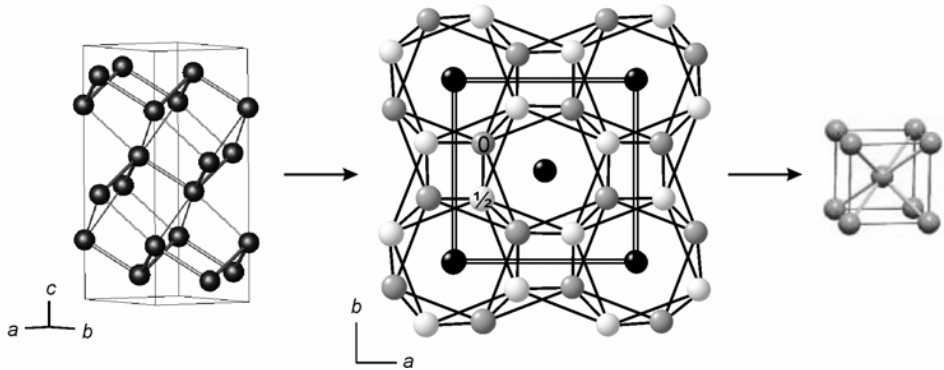

*Figure 2.* The crystal structure of group-V elements: (left) As-type ($hR2$), (middle) host-guest structure, and (right) bcc

### 2.2.2. *Phosphorus*

The lighter group-V element phosphorus forms the As-type (*hR*2) structure only under pressure, above 5 GPa (Table 2), and at 10 GPa transforms to the simple cubic structure, similar to arsenic. Despite of its low packing density, the simple cubic structure is stable in phosphorus over a very wide pressure range up to 103 GPa. It has been long know that on further pressure increase, the simple cubic structure transforms to a simple hexagonal structure via a complex phase (*Akahama et al. 1999*) that has remained unsolved until recently. Amongst several suggestions, a host-guest structure of Ba-IV type (similar to Bi-III) has been proposed as a candidate for this intermediate phase from the theoretical calculations (*Ehlers and Christensen 2004*). We know, however, that the coordination number of the Bi-III host-guest structure is 9-10, which is too high for an intermediate phase between simple cubic (coordination number 6) and simple hexagonal phase (coordination number 8). It is only recently that metadynamic calculations (*Ishikawa et al. 2006*) revealed a modulated atomic structure, while the subsequent experiment demonstrated an incommensurate nature of the modulation (*Fujihisa et al. 2007*) with a coordination number equal to 6. The simple hexagonal structure is shown to transform to bcc at 260 GPa (*Akahama et al. 2000*).

### 2.3. GROUP-VI ELEMENTS

Elemental sulphur, a yellow insulating mineral, is composed of crown-like 8-member ring molecules that form a complex crystal structure. Compression has a dramatic effect on these molecules: they break apart forming various chain and ring structures on pressure increase (*Degtyareva et al. 2007a* and references therein). Sulphur becomes metallic and superconducting above 90 GPa with $T_c = 10$ K (*Struzhkin et al. 1997*) forming an incommensurately modulated crystal structure at 300 K (Figure 3) (*Degtyareva et al. 2005, Hejny et al. 2005*). The structure is described by a (3+1)-dimensional superspace group $I'2/m(0\beta0)s0$, and is the same as found earlier in the heavier group-VI elements Se and Te (see *McMahon and Nelmes 2006* and references therein).

This incommensurately modulated structure forms on pressure increase between the chain structures of S, Se and Te with the 2 nearest neighbors (according to the 8-N coordination rule) and the β-Po structure (found at high temperature in Te and at room temperature in Se and S) with the coordination number 6+2. In Se and Te, the β -Po structure is known to transform to bcc on pressure increase with coordination number 8+6 (see *McMahon and Nelmes*

*2006* and references therein). The interatomic distances in the modulated structure are also modulated and give an effective coordination number of 6 (2+2+2), thus being an intermediate step in the structural sequence from ring and chain arrangements to typical metallic structures that shows an increase in coordination number and packing density. For the analysis of the pressure effect on the interatomic distances and coordination number in Se, see (*Degtyareva et al. 2005*).

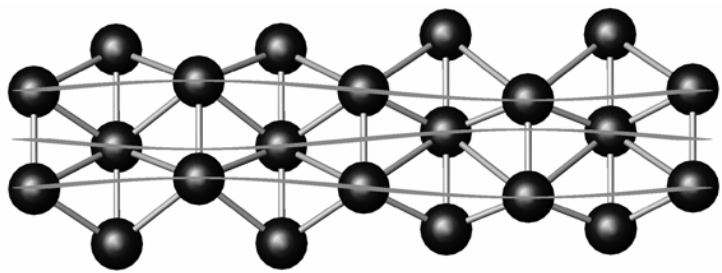

*Figure 3.* Crystal structure of the high-pressure phase of sulfur, S-IV, that is incommensurately modulated (commensurate approximant is shown in projection down the monoclinic a-axis) (from *Degtyareva et al. 2007b*)

## 2.4.  GROUP-I ELEMENTS (ALKALI METALS)

On the example of the transition sequences of the group IV, V, and VI elements, we have learned that the covalent and molecular structures respond to high pressure by increasing the coordination number and packing density, transforming to metallic state. Complex crystal structures appear on pressure increase as an intermediate step toward the closely-packed structures. Now to the question "what structures are expected in elements under high pressure" one could intuitively answer "the closely-packed structures". However, the experimental studies on alkali and alkali-earth metals have shown that for them the close-packed structure is not close enough. Now we look at the "non-simple behaviour of simple metals at high pressure" (*Maksimov et al. 2005*).

### 2.4.1.  *Results from experimental work*

The alkali metals are considered as textbook examples of nearly-free electron metals at ambient conditions. They crystallize in a highly symmetric and closely-packed body-centred cubic (bcc) structure. On pressure increase, the bcc structure transforms to a close-packed arrangement of the face-centred cubic (fcc) structure, and then to unexpectedly complex low-symmetry structures (Figure 4) (see reviews *Degtyareva 2006, McMahon and Nelmes 2006, Schwarz 2004, Syassen 2002*).

TABLE 3. Sequences of structural transformations on pressure increase for group I elements. The numbers above the arrows show the transition pressures in GPa. "H-g" stands for a host-guest structure, "dhcp" for a double hexagonal close-packed structure

| **Li** | 7.5          39          42<br>bcc $\rightarrow$ fcc $\rightarrow$ $hR1$ $\rightarrow$ $cI$ 16 < 50GPa |
|--------|---------------------------------------------------------------------------------------------------------|
| **Na** | 65        104        117        125        180<br>bcc $\rightarrow$ fcc $\rightarrow$ $cI$ 16 $\rightarrow$ $oP8$ $\rightarrow$ h-g $\rightarrow$ $hP4$ < 200 GPa |
| **K**  | 11.6        20          54        90        96<br>bcc $\rightarrow$ fcc $\rightarrow$        h-g $\rightarrow$ $oP8 \rightarrow tI4$ $\rightarrow$ $oC16$ < 112 GPa |
| **Rb** | 7          13          17        20        48<br>bcc $\rightarrow$ fcc $\rightarrow$ $oC58$ $\rightarrow$ h-g $\rightarrow$ $tI$ 4 $\rightarrow$ $oC16$ < 70 GPa |
| **Cs** | 2.4      4.2          4.3        12        72<br>bcc $\rightarrow$ fcc $\rightarrow$ $oC84$ $\rightarrow$        $tI$ 4 $\rightarrow$ $oC16 \rightarrow$ dhcp < 223 GPa |

Lithium and sodium adopt a complex cubic structure with 16 atoms in the unit cell (Pearson symbol $cI$16) at around 40 GPa (*Hanfland et al. 2000*) and at 104 GPa (*McMahon et al. 2007b, Syassen 2002*), respectively (Table 3). On further compression the $cI$16 phase of Na transforms at 118 GPa to an orthorhombic structure with 8 atoms in the unit cell ($oP8$) (*Gregoryanz et al. 2008*), a novel structure-type for an element. At 125 GPa, the $oP8$ phase transforms to an incommensurate host-guest structure (*Gregoryanz et al. 2008, Lundegaard et al. 2009a*) with the same 16-atom tetragonal host structure observed previously in K and Rb (*McMahon et al. 2001 & 2006a*). Melting line of Na showed a peculiar behavior (*Gregoryanz et al. 2005*) with a bend-over at 31 GPa and 1000 K that reached its minimum of room temperature at 118 GPa. Numerous complex phases with large unit cells were discovered in the vicinity of the melting line minimum (*Gregoryanz et al. 2008*). The most recent experimental work on Na reported quite an astonishing result that above 180 GPa sodium becomes optically transparent and insulating; complementary theoretical calculations proposed for this phase a hexagonal structure with four atoms in the unit cell, $hP4$ (*Ma et al. 2009*). At the same time, a publication appeared on a pressure-induced metal to semiconductor transition in Li (*Matsuoka and Shimizu 2009*) at pressures near 80 GPa, the experimental determination of the corresponding crystal structures however has not yet been reported.

Heavier alkali metals, K, Rb, and Cs, have been long known to assume complex crystal structures under pressure above the stability region of fcc, at much lower pressures than in Li and Na (Figure 4, Table 3). These include orthorhombic structures $oC52$ in Rb and $oC84$ in Cs, a host-guest structure in K and Rb, tetragonal $tI4$ and an orthorhombic $oC16$ structures in Rb and Cs. A very recent high-pressure study on K reported an observation of an $oP8$ structure, similar to that in Na, and the $tI4$ and $oC16$ structures, same as known for Rb and Cs (*Lundegaard et al. 2009b*).

The phase transitions from fcc to complex structures in all alkali metals are accompanied by a decrease in symmetry, coordination number and packing density. A subsequent increase in coordination number is observed at the transition to the $oC16$ in Rb and Cs and to the double hexagonal close-packed (dhcp) structure in Cs.

### 2.4.2.  *Results from theoretical work*

The appearance of some of the complex structures in the heaviest alkali metals Rb and Cs is suggested to be connected to an electronic *s*-to-*d* transfer (see review *Maksimov et al. 2005* and references therein). The transition from fcc to complex phases at high pressures in Li and Na is explained with an *s*-to-*p* band overlap (*Christensen & Novikov 2001, Hanfland et al. 2000*). However, the observation of a large number of structural transformations and a variety of structural types in alkali metals under pressure require consideration of other factors that might be responsible for the formation of these phases. Hume-Rothery effect has been suggested as a stabilization mechanism for the $cI16$, $oC52$ and $oC84$ structures (*Degtyareva 2006*). This model explains the appearance of the complex structures as a result of lowering the electronic energy and thus the total energy of the crystal structure due to a formation of additional planes in a Brillouin-Jones zone in a close contact with the Fermi surface. The electronic contribution into total energy becomes more important on compression, adding significantly to the stabilization of the $cI16$ and related phases.

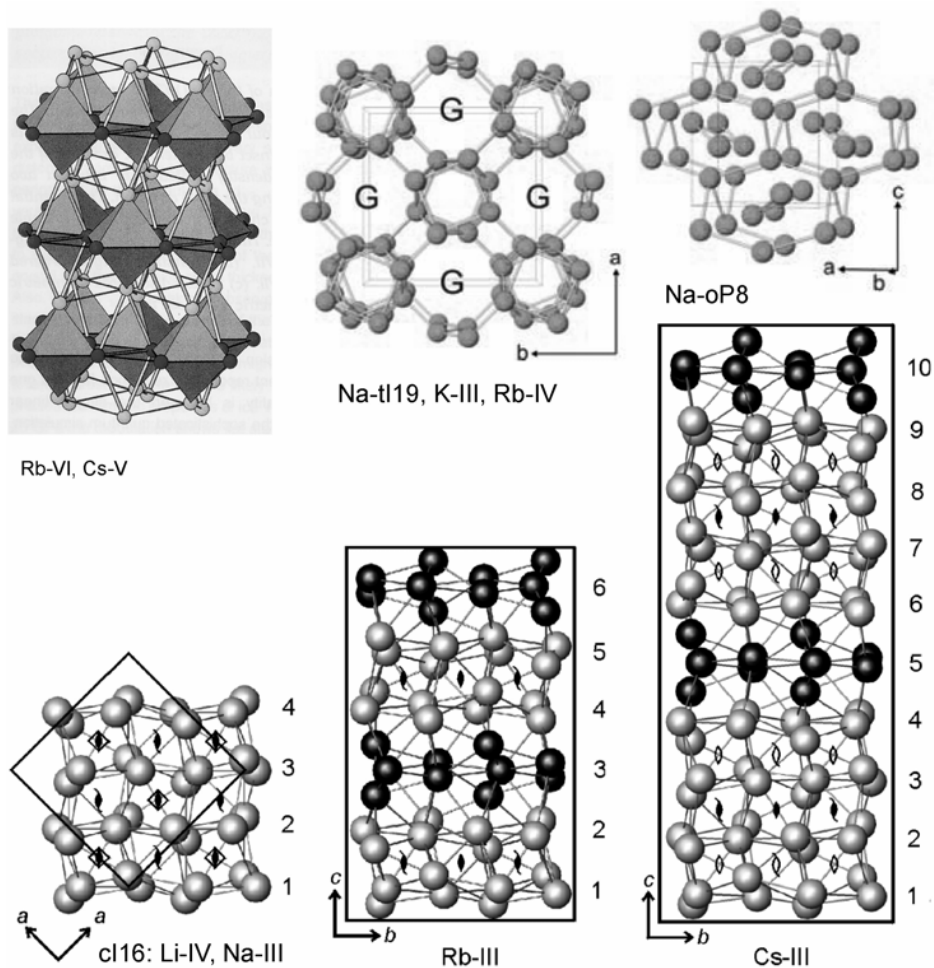

*Figure 4.* Complex crystal structures of alkali metals forming at high pressure (from *Degtyareva 2003, Gregoryanz et al. 2008, Schwarz et al. 1998*). For the host-guest structure of Na, K and Rb, "G" marks channels occupied by chains of guest atoms, different arrangements are found in different elements (see *Lundegaard et al. 2009a, McMahon et al. 2001 & 2006a*)

At higher pressures in the post-*cI*16 region, the density of Li and Na is so high ($\rho/\rho_0 \approx 4$ for Na) that the atomic cores are beginning to touch (*Katsnelson et al. 2000, Neaton and Ashcroft 1999, 2001,*) and new physical phenomena come into play. For both Li and Na, theoretical calculations find that electrons localize in the interstitial regions of the structures (*Christensen and Novikov 2001, Ma et al. 2008, Neaton and Ashcroft 1999, 2001, Rousseau et al. 2005,*

*Zhou et al. 2009*). These computational studies have suggested several crystal structures for the post-*cI*16 regime in Li and Na, with the most recent reports (*Pickard & Needs 2009, Yao et al. 2009a*) predicting insulating and low-coordinated structures for dense Li containing 24 atoms in the unit cell. The structural behavior uncovered by experimental work, however, turned out to be much more complex than the prediction, with crystal structures containing a very large number of atoms in the unit cell (*Gregoryanz et al. 2008 & 2009*). There is an alternative view suggesting that the dense phases of alkali metals undergo partial core ionization and attain valence greater than one as, for example, described for *tI*4 and *oC*16 structures of Rb and Cs and *oP*8 structure of Na (*Degtyareva 2006, Degtyareva and Degtyareva 2009*).

## 2.5. GROUP-II ELEMENTS (ALKALI-EARTH METALS)

The alkali-earth metals, Ca and Sr, crystallize at ambient conditions in the close-packed face-centered cubic (fcc) structure, both transforming to the body-centered cubic (bcc) structure under pressure. The heavier group-II element Ba has the bcc structure already at ambient pressure at room temperature forming the fcc phase at high temperatures. These three elements showed a surprising and puzzling behavior under pressure, transforming from closely-packed to low-coordinated structures with a reduction of packing density – not the behavior one would intuitively expect on compression.

First we discuss cubic structure and peculiar Ca-IV and Ca-V phases. Then we talk about Sr and Ba and their host-guest structure which discovery marked a new era of incommensurate crystal structures of elements.

TABLE 4. Sequences of structural transformations on pressure increase for group II elements. "H-g" stands for a host-guest structure. The numbers above the arrows show the transition pressures in GPa

| **Ca** | $\xrightarrow{20}$ bcc $\xrightarrow{32}$ *cP*1 $\xrightarrow{113}$ *tP*8 $\xrightarrow{139}$ *oC*8 < 154 GPa |
|---|---|
| **Sr** | fcc $\xrightarrow{3.5}$ bcc $\xrightarrow{27}$ β-Sn, *tI*4 $\xrightarrow{35}$ *mI*12 $\xrightarrow{49}$ h-g <117 GPa |
| **Ba** | bcc $\xrightarrow{5.5}$ hcp $\xrightarrow{12}$ h-g $\xrightarrow{45}$ hcp < 105 GPa |

### 2.5.1. *Calcium*

Calcium transforms from fcc to bcc at 20 GPa, and at higher pressure of 32 GPa to a simple cubic structure (sc, *cP*1) (*Olijnyk and Holzapfel 1984*). The transition from bcc to sc in Ca occurs with a decrease in packing density and coordination number, which changes from 14 (=8+6) in bcc down to 6 in sc phase. The simple cubic structure is stable in Ca in a very large pressure range from 32 to 113 GPa (Table 4). Above this pressure, complex structures are observed: Ca IV stable between 113 and 139 GPa and Ca-V stable above this pressure. Their crystal structures have only recently been identified experimentally: Ca-IV has a primitive orthorhombic structure with 8 atoms in the unit cell (*oP*8), space group $P4_12_12$; Ca-V has a C-centered orthorhombic structure with 8 atoms in the unit cell (*oC*8), space group *Cmca* (*Fujihisa et al. 2008*). They were confirmed to be identical to the theoretical models that were predicted recently (*Ishikawa et al. 2008a*). The structures of Ca-IV and V have a complex atomic coordination with 7 nearest neighbors. These structures have never been seen before in other alkaline-earth metals or in other elements.

Interestingly, the simple cubic phase in calcium presents serious theoretical problems. For instance, recent calculations revealed that it is mechanically unstable in the corresponding pressure range (*Gao et al. 2008, Teweldeberhan and Bonev 2008*), prompting further calculations that tried to reconcile the theory and experiment (*Errea et al. 2008*). The most recent experimental study on Ca was devoted to checking the stability of the simple cubic structure (*Gu et al. 2009*). Subsequent resent theoretical work found that the simple cubic phase is indeed stable if calculated at 300K, while another structure is preferred at lower temperatures (*Yao et al. 2009b*). Generally, phase transitions in alkali-earth metals are attributed to an *s* to *d* electronic transfer (see review *Maksimov et al. 2005*), with an alternative explanation suggesting that Ca in the simple cubic structure attains a valence greater than two (*Degtyareva 2006*).

### 2.5.2. *Strontium and Barium*

The discovery of the host-guest structure in Ba in 1999 (*Nelmes et al. 1999*) and subsequently in Sr (*McMahon et al. 2000b*) marked a beginning of a new era in the high-pressure crystallography of elements. This type of structure has not been observed before in any other element and has not been expected either. The finding was quite astonishing and hard to comprehend but now the existence of the host-guest structure in elemental metals is generally accepted and it is a widely spread phenomenon at high-pressure.

Between 35 and 49 GPa before the transformation to the host-guest structure, strontium shows a unique and peculiar structure (Sr-IV). It is body-centered monoclinic, space group *Ia*, with 12 atoms in the unit cell (Pearson notation *mI*12), and can be viewed as a helical distortion of the tetragonal β-Sn structure (Sr-III), in which atoms lying in (straight) chains along the Sr-III c-axis are displaced to form helical chains (*Bovornratanaraks et al. 2006*). The existence of the β-Sn structure in Sr in the pressure range 26-35 GPa is very peculiar, it has never been found in any other alkaline-earth metal, but is known to be a stable phase of group-IV elements (see Section 2.1). The stability of the β-Sn in Sr received a very limited attention, and similar to the simple cubic structure of Ca, presents difficulties for theoretical calculations (*Phusittrakool et al. 2008*).

Analyzing the changes in coordination number in Sr under pressure, we can see a dramatic decrease from 14 (=8+6) in bcc down to 6 (=2+4) in β-Sn and *mI*12 structures. There is a reversed increase in coordination number at the following transition to the host-guest structure with further pressure increase: the coordination number is 9-10 (this number is given in analogy with the Bi-III host-guest structure, however single crystal structure refinement of atomic modulations is required for a determination of the exact number). Similar reversed increase in coordination number was shown for alkali metals Rb and Cs.

## 3.   Some remaining questions and summary

### 3.1.  GENERAL POINTS

Simple elements exhibit numerous phase transformations under pressure, forming complex crystal structures, often incommensurate, also called aperiodic. Incommensurate structures found in elements up until the year 2004 are summarized in (*McMahon and Nelmes 2004a*) with subsequent findings described in the review (*McMahon and Nelmes 2006*). A higher-dimensional formalism is needed for the crystallographic description of the incommensurate structures of elements. Explanation of the meaning of the superspace group symbols can be found, for example, in (*Janssen et al. 2004*). A detailed crystallographic description of the host-guest structures in group-V elements is presented in (*Degtyareva et al. 2004b*), while general higher-dimensional superspace formalism is outlined in (*van Smaalen 1995, van Smaalen 2007*). Refs. (*van Smaalen 2004*) and (*Petrıcek and Dusek 2004*) give an elementary introduction to superspace crystallography with some examples. A very detailed and accessible description "for pedestrians" of this crystallographic formalism applied to host-guest (composite) structures is given in (*Sun et al. 2007*).

The numerous structural transitions in simple elements described above are accompanied by marked changes in the physical properties. As already mentioned, initially insulating sulphur becomes superconducting at around 90 GPa (*Struzhkin et al. 1997*), while sodium, initially a free-electron metal, turns into optically transparent insulator at around 200 GPa (*Ma et al. 2009*). Calcium becomes an element with the highest *T*c among all pressure-induced superconductors, with the critical temperature reaching 25K at 161 GPa in the complex Ca-V phase (*Yabuuchi et al. 2006*). The correlation between the resistivity, superconductivity and crystal structure in simple elements is discussed in the review (*Degtyareva 2006*).

## 3.2. HOST-GUEST STRUCTURES

The discovery of incommensurate host-guest structures in elemental metals under pressure has demonstrated unexpected structural complexity in compressed solids, and has challenged theories of condensed matter. First found in the group-II elements Ba and Sr (*McMahon et al. 2000b, Nelmes et al. 1999*), the host–guest structure is also shown to form at high pressures in the group-I elements Rb (*McMahon et al. 2001*), K (*McMahon et al. 2006a*), and Na (*Lundegaard et al. 2009a*) the group-V elements Bi, Sb and As (*Degtyareva et al. 2004 a & b, McMahon et al. 2000a, Schwarz et al. 2003,*) and in the transition metal Sc (*McMahon et al. 2006b*) Interestingly, the same host structure is found for the group-II and group-V elements and the transition metal Sc, while another type of host structure is observed for group-I elements. Guest structure in all elements forms chains of atoms that are correlated in various crystal structures or show a disorder between chains.

In the group-V elements, the host-guest structure plays a role of an intermediate step between the semi-metallic open packed and metallic closely-packed structures, while in group-I and II elements a more intricate picture of changes in electronic structure emerges. Such complex atomic arrangement provides a minimization of the electronic energy reducing the overall structural energy.

Several interesting phenomena have been observed in these elemental host-guest structures since their discovery in 1999, including atomic modulation of the host and guest components (*McMahon et al. 2007a*), incommensurate-to-incommensurate phase transitions (*Degtyareva et al. 2004a*), and 'melting' of the guest chains (*McMahon and Nelmes 2004b*). An effect of site ordering has been studied on Bi-Sb binary alloys (*Haussermann et al. 2004*), which showed a formation of a site-disordered incommensurate host-guest structure. First-principles calculations have been performed on the host-guest structures of group II and V elements (*Reed and Ackland 2000, Haussermann et al. 2002, Ormeci and Rosner 2004*) and more recently of the group-I element Na (*Ma et al. 2009, Lazicki et al. 2009*) and transition metal Sc (*Ormeci et al. 2006, Arapan et al. 2009*) using commensurate approximants to describe their electronic structure. For the group-V elements, these calculations revealed that the electronic structure of the host-guest phases is intermediate between that of the low-pressure semi-metallic phases and the high-pressure metallic bcc phase (*Haussermann et al. 2002*), in accord with our conclusions of Section 2.2.1 based on the analysis of the coordination number.

To better understand their properties, experimental techniques other than X-ray diffraction are being applied to the high-pressure host-guest structures of elements, such as Raman scattering (*Degtyareva et al. 2007c, Wang et al. 2006*) and inelastic X-ray scattering (*Loa et al. 2007*).

## 3.3. INCOMMENSURATELY MODULATED STRUCTURES

Pressure-induced incommesurately modulate structures have been discovered in elements in the year 2003 simultaneously and independently in the high-pressure phases of group-VI element Te and group-VII element iodine (see review *McMahon and Nelmes 2006* and references therein). Since then, the incommensurately modulated structures have been found in Se, S and P and suggested in Br, displaying great structural diversity. The group VII element iodine (and probably bromine) form a $Fmm2(\alpha00)0s0$ modulated structure (*Kume et al. 2005, Takemura et al. 2003*), the group-VI elements Te, Se and S adopt $I'2/m(0\beta0)s0$ structure (*Hejny and McMahon 2003, McMahon and Nelmes 2006*), the group-V element phosphorus has a $Cmmm(00\gamma)s00$ modulated structure (*Fujihisa et al. 2007*). Additionally, a long-period structure of a high-pressure phase of Ga, Ga-II (*Degtyareva et al. 2004c*), was interpreted as a commensurately modulated $Fddd(00\gamma)0s0$ structure (*Perez-Mato et al. 2006*). At ambient pressure, a modulated structure is known only in one element: uranium.

Theoretical calculations were performed on the modulated structures of phosphorus, tellurium, selenium, sulphur, iodine and bromine, using commensurate approximants, to study the electronic origins of the formation of these structures (*Ackland and Fox 2005, Degtyareva et al. 2007b, Duan et al 2007, Ishikawa et al. 2008b, Loa et al. 2009, Marqués et al. 2008, Nishikawa 2007*). A charge-density wave instability is found to be responsible for the structural modulation in sulphur, in competition with the superconducting state (*Degtyareva et al. 2007b*). An inelastic X-ray scattering technique applied to the incommensurately modulated phase of Te found a phonon softening associated with the structural modulation, while complementary theoretical calculations found a Fermi surface nesting within this phase (*Loa et al. 2009*).

## 4.   Conclusions

The discussed phase transitions in the simple elements revealed the following trends. On pressure increase, the group IV, V, VI and VII elements undergo phase transformations from non-metallic open-packed to metallic closely packed structures with an increase in coordination number and packing density, passing through some very complex atomic arrangements including incommensurate structures. The group I and II elements, the alkali and alkali-earth metals, that form closely-packed structures at ambient pressure, reveal surprisingly complex behavior under pressure, exhibiting structures with very large number of atoms in the unit cell owing it to the complicated changes in their electronic structure.

### ACKNOWLEDGEMENTS

The support from the Royal Society is greatly acknowledged. The author would like to thank Valentina Degtyareva for fruitful discussions.